\documentclass[jkps,preprint,fleqn,showpacs,showkeys,twocolumn,10pt]{revtex4}
\usepackage{graphicx}
\usepackage{amssymb}
\usepackage{amsmath}
\usepackage{bm}
\begin{document}
\setcounter{page}{0}
\title[]{On the distribution of dark matter in galaxies: quantum treatments}
\author{Carlos R. \surname{Arg\"uelles}}
\email{carlos.arguelles@icranet.org}
\author{Remo \surname{Ruffini}}
\affiliation{Dipartimento di Fisica and ICRA, Sapienza Universit\`a di Roma, P.le Aldo Moro 5, I--00185 Rome, Italy}
\affiliation{ICRANet, P.zza della Repubblica 10, I--65122 Pescara, Italy}
\author{Ivan \surname{Siutsou}}
\affiliation{Dipartimento di Fisica and ICRA, Sapienza Universit\`a di Roma, P.le Aldo Moro 5, I--00185 Rome, Italy}
\affiliation{ICRANet, P.zza della Repubblica 10, I--65122 Pescara, Italy}
\author{Bernardo \surname{Fraga}
}
\affiliation{Dipartimento di Fisica and ICRA, Sapienza Universit\`a di Roma, P.le Aldo Moro 5, I--00185 Rome, Italy}
\affiliation{Universit\'e de Nice Sophia di Antipolis, Nice, France}

\date[]{}

\begin{abstract}
The problem of modeling the distribution of dark matter in galaxies in terms of equilibrium configurations of collisionless self-gravitating quantum particles is considered. We first summarize the pioneering model of a Newtonian self-gravitating Fermi gas in thermodynamic equilibrium developed by Ruffini and Stella (1983), which is shown to be the generalization of the King model for fermions. We further review the extension of the former model developed by Gao, Merafina and Ruffini (1990), done for any degree of fermion degeneracy at the center ($\theta_0$), within general relativity. Finally, we present here for the first time the solutions of the density profiles and rotation curves corresponding to the Gao et. al. model, which have a definite mass $M_h$ and circular velocity $v_h$, at the halo radius $r_h$ of the configurations, typical of spiral galaxies. This treatment allow us to determine a novel core-halo morphology for the dark matter profiles, as well as a novel particle mass bound associated with them.
\end{abstract}



\maketitle

\section{INTRODUCTION}
One of the most quoted papers in the study of dark matter distribution in galactic halos is certainly the work of Tremaine \& Gunn \cite{PRL79Tremaine}, there the authors established a lower limit on the mass of neutrinos composing galactic halos, by considering an isothermal classical distribution of self-gravitating particles, and imposing quantum constraints on the phase space density in the core of the galaxies. This treatment presenting a peculiar mixture of quantum and classical considerations has always attracted the attention and the suspicion of many astrophysicists and theoretical physicists. It was in particular the opinion of one of us (R. Ruffini), that a self-consistent treatment of quantum constraints in a self-consistent quantum description of the microphysical system was needed. A long effort so started.

\section{HISTORICAL REVIEW: TWO PIONEERING WORKS}
One of the preliminary works in order to prove the analogy and differences between a classical and quantum self-gravitating system was advanced within a Newtonian approach in Ruffini \& Stella (1983) \cite{AA83Ruffini}. There, the problem of a semi-degenerate system of fermions under gravitational interaction was approached, and compared and contrasted with the classical King model.

They proposed a distribution function built for non-relativistic particles, which reads:
\begin{align}%\label{}
f(v)&=&\frac{1-\exp{[-j^2(v_e^2-v^2)]}}{\exp{[j^2(v^2-\bar{\mu})]}+1},\, v\leq v_e \nonumber\\
    &=&0, \qquad \qquad \qquad \qquad \quad v>v_e,\nonumber
%\label{eq:1}
\end{align}
where $v_e$ is the escape velocity. In the limit $v_e\rightarrow\infty$, the usual Fermi distribution is obtained. The other two constant parameters are $j^2=m/(2kT)$ and $\bar{\mu}=2\mu/m$. The relevance of this $f(v)$ is that it is an extension of the King model to the case of a Fermi gas. Moreover, if the degeneracy parameter $\theta=j^2\bar{\mu}$ is defined, it can be easily seen that when $\theta\rightarrow-\infty$, the non-degenerate limit is reached and the distribution function used by King is recovered. Instead, when $j^2\rightarrow\infty$ and $\bar{\mu}\rightarrow v_e^2$ the degenerate limit is obtained, and the escape velocity is associated with the Fermi energy.

The energy integral,
\begin{equation}
E=v^2/2+V(r)\, ,
\end{equation}
together with the Jeans theorem for spherical systems, allowed them to simply relate the escape velocity with the gravitational potential by $v_e^2=-2V$ (being $V=0$ at the surface of the configuration). They finally solved the Poisson equation for $W=-2j^2V$, being the mass density given by $\rho\propto\int f(v)v^2dv$ which is related with $W$ via the $j$ parameter.

\begin{figure}[!hbtp]
\centering
\includegraphics[width=1.\hsize,clip]{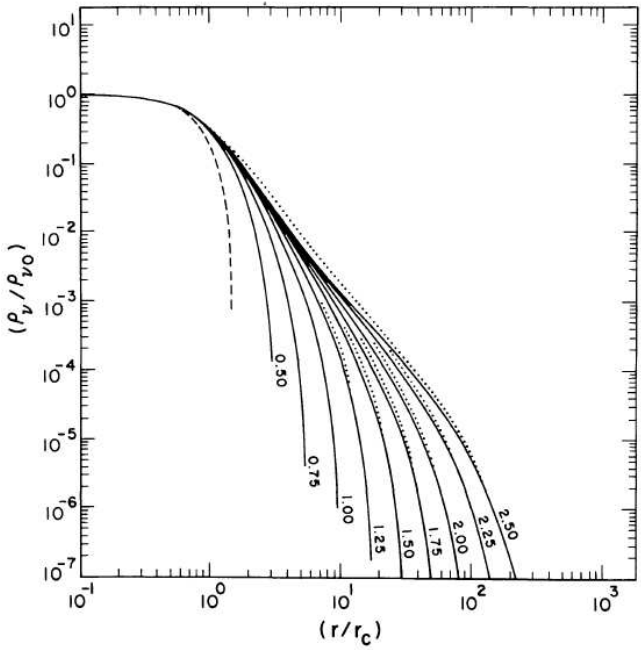}
\caption{Normalized density profiles for different values of $W_0$ and fixed $\theta_0=0$. The dotted curve corresponds to the analogous King profiles while the dashed curve represents the degenerate limit (taken from \cite{AA83Ruffini} with permission).  }\label{fig:1}
\end{figure}

For simplicity, in the attempt to understand the physical interpretation of the parameters, an only value for the central degeneracy parameter was assumed for the sake of example,
\begin{equation}
\theta(0)\equiv\theta_0=0\, ,
\end{equation}
and no other values for $\theta_0$ were explored at the time. The different normalized mass density solutions were given for different values of $W(0)\equiv W_0$ as shown in Fig.~\ref{fig:1}.

Under these special conditions the analogy between the self-gravitating system of fermions and the King model was proven, as well as a first attempt to justify the Tremaine \& Gunn limit.

It soon became clear that these solutions, although interesting in reproducing classical results of the King profiles for a self-gravitating fermion gas, were really extremely restrictive, and not representative of the general solutions for a relativistic self-gravitating system of massive fermions. These restrictions correspond to three different constraints: 1) $\theta_0=0$; 2) the application of a cut-off in the phase space which implies the elimination of an important family of solutions; and 3) the use of a non relativistic Newtonian approach.

A fundamental step was made by Gao, Merafina and Ruffini (1990) \cite{AA90Gao}, to include special relativity effects in the phase space of the distribution function, as well as general relativity. Thus, they considered the relativistic Fermi-Dirac distribution function for the `inos' \cite{footnote} without any cut-off in their momentum space, i.e. $f(\epsilon)=(\exp[(\epsilon-\mu)/(k T)]+1)^{-1}$, where $\epsilon(p)=\sqrt{c^2 p^2+m^2 c^4}-mc^2$ is the particle kinetic energy and $\mu$ the chemical potential with the particle rest-energy subtracted off.

They wrote the system of Einstein equations in the spherically symmetric metric $g_{\mu \nu}={\rm diag}(e^{\nu},-e^{\lambda},-r^2,-r^2\sin^2\theta)$,
where $\nu$ and $\lambda$ depend only on the radial coordinate $r$, together with the thermodynamic equilibrium conditions of Tolman \cite{PR30Tolman}, and Klein \cite{RMP49Klein},
\begin{equation}
e^{\nu/2} T=const.\, , \quad e^{\nu/2}(\mu+m c^2)=const. \nonumber
\end{equation}
in the following dimensionless way,
\begin{align}
		&\frac{d\hat M}{d\hat r}=4\pi\hat r^2\hat\rho \label{eq:1}\\
		&\frac{d\theta}{d\hat r}=\frac{\beta_0(\theta-\theta_0)-1}{\beta_0}
    \frac{\hat M+4\pi\hat P\hat r^3}{\hat r^2(1-2\hat M/\hat r)}\\
    &\frac{d\nu}{d\hat r}=\frac{\hat M+4\pi\hat P\hat r^3}{\hat r^2(1-2\hat M/\hat r)} \\
    &\beta_0=\beta(r) e^{\frac{\nu(r)-\nu_0}{2}}\, . \label{eq:2}
\end{align}
There, the following dimensionless quantities were introduced: $\hat r=r/\chi$, $\hat M=G M/(c^2\chi)$, $\hat\rho=G \chi^2\rho/c^2$ and $\hat P=G \chi^2 P/c^4$, where $\chi=2\pi^{3/2}(\hbar/mc)(m_p/m)$ is the dimensional factor which has unit of length and scales as $m^{-2}$; with $m_p=\sqrt{\hbar c/G}$ the Planck mass, and the temperature and degeneracy parameters, $\beta=k T/(m c^2)$ and $\theta=\mu/(k T)$, respectively. The mass density $\rho$ and pressure $P$ are expressed in terms of the standard infinite integrals in momentum space weighted with the $f(\epsilon)$ already given, for a relativistic and semi-degenerate Fermi gas (see \cite{AA90Gao}).

\begin{figure}[!h]
\centering
\includegraphics[width=1.\hsize,clip]{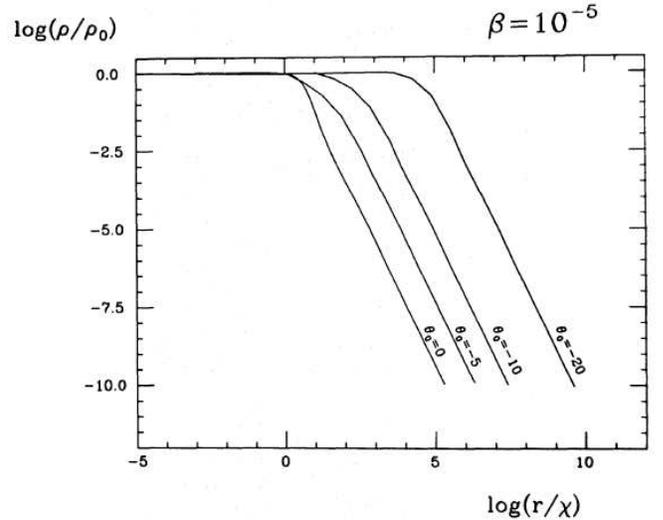}
\caption{Different density profiles for different $\theta_0<0$ and fixed $\beta_0$ in dimensionless variables. To note the simple cored plus $r^{-2}$ morphology (taken from \cite{AA90Gao} with permission).}\label{fig:2}
\end{figure}

\begin{figure}[!h]
\centering
\includegraphics[width=1.\hsize,clip]{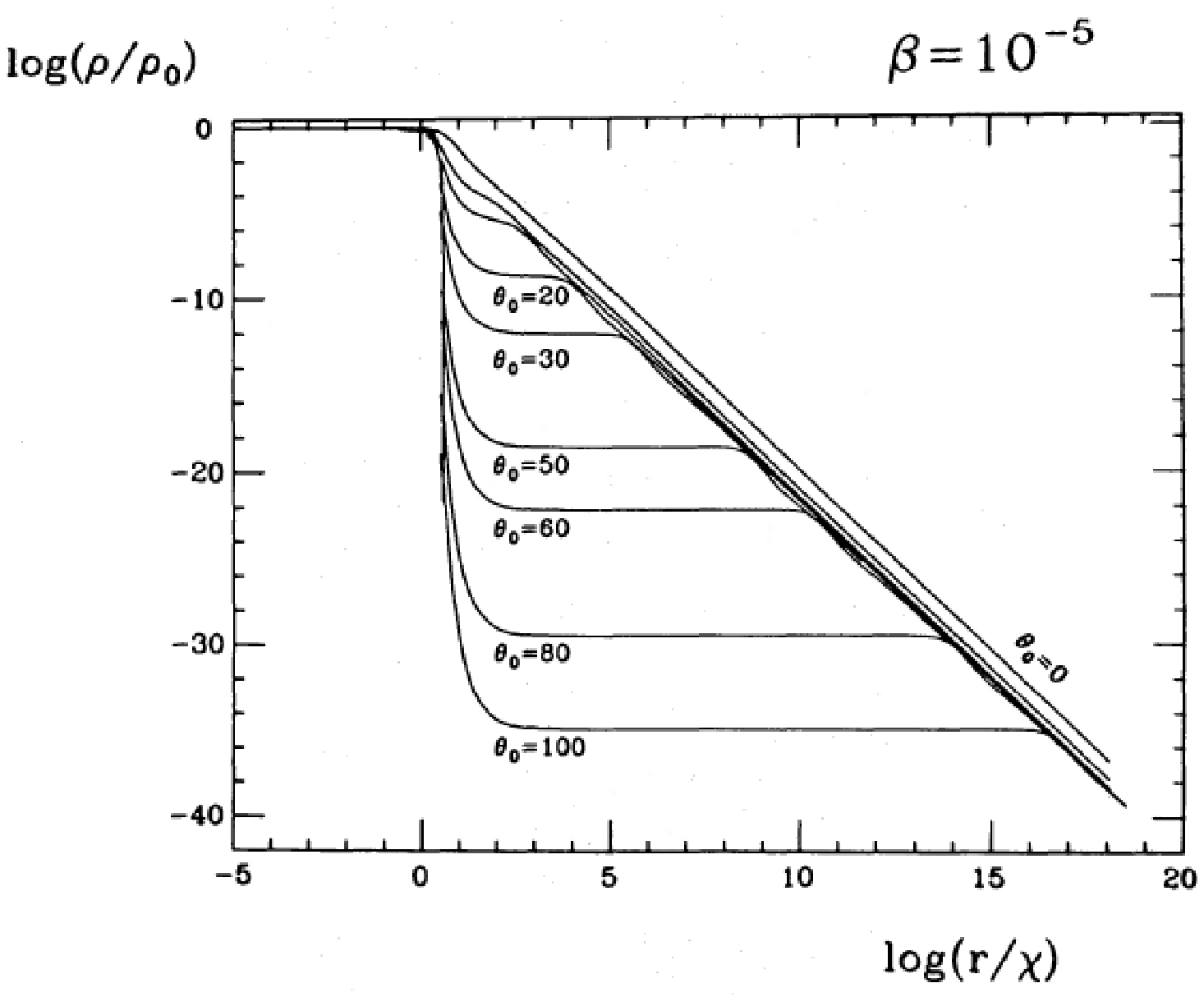}
\caption{Different density profiles for different $\theta_0\geq0$ and fixed $\beta_0$ in dimensionless variables. To note the more complex core plus `plateau' plus $r^{-2}$ morphology (taken from \cite{AA90Gao} with permission).}\label{fig:3}
\end{figure}

In that work, they solved the initial condition problem for the variables of the system, $\theta(r)$, $\beta(r)$, $\nu(r)$, and $M(r)$, by giving at $r=0$ (and indicated by a subscript `0') $M_0=0$, while giving arbitrary values for the temperature and degeneracy parameters $\beta_0$ and $\theta_0$, respectively. In Figs.~\ref{fig:2}--\ref{fig:3}, different normalized mass density solutions for different $\theta_0<0$ and $\theta_0\geq0$ respectively are shown, for a fixed non-relativistic central temperature parameter $\beta_0$.

It is important to note that the system (\ref{eq:1}--\ref{eq:2}) has no particle mass $m$ dependence when solved in the dimensionless variables, while instead the physical magnitudes such as $r$ and $\rho$ have an explicit dependence on $m$ trough the dimensional factor $\chi(m)$. The fact that they were mainly interested in the general properties of the solutions without going through the physical magnitudes, no particle mass constraints were put there.

\section{NOVEL PROCEDURE AND DISCUSSION}
We have recently returned to the Gao et. al. work, and propose a completely different way for solving the boundary condition problem for the system (\ref{eq:1}--\ref{eq:2}), in order to fulfill the observationally inferred values of typical dark matter halos in spiral galaxies as given in \cite{AJ08deBlok}. Namely, for given initial conditions $M_0=\nu_0=0$, arbitrary $\theta_0$ (depending on the chosen central degeneracy), and defining the halo radius $r_h$ at the onset of the flat rotation curve, we solve an eigenvalue problem for the central temperature parameter $\beta_0$, until the observed halo circular velocity $v_h$ is obtained. After this, we solve a second eigenvalue problem for the particle mass $m$ until the \textit{observed} halo mass $M_h$ is reached at the radius $r_h$.

The quest has been to use all these information in order to put constraints on the mass of the `ino' in galactic halos by introducing the observational properties possibly to be utilized in this research.

\begin{figure}[!hbtp]
\centering
\includegraphics[width=1.\hsize,clip]{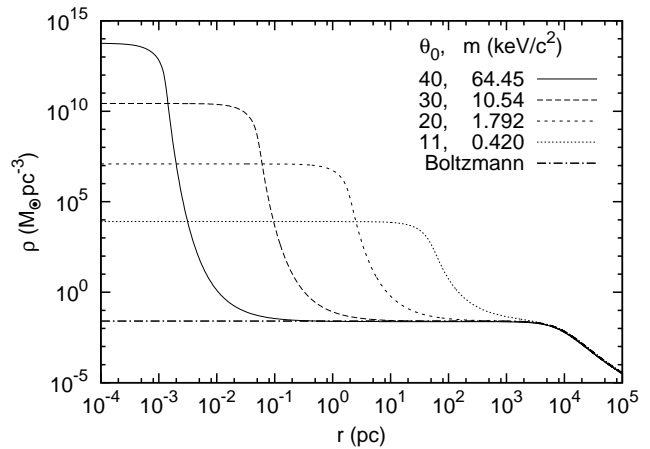}
\caption{Physical density profiles for specific ino masses $m$ and central degeneracies $\theta_0$ fulfilling the observational constraints $M_h=1.6\times10^{11} M_\odot$ and $v_h=168$ km/s at $r_h=25$ Kpc (as taken from \cite{AJ08deBlok} and detailed in \cite{PRL13Ruffini}). In dot-dashed line the purely Boltzmannian profile for comparison.}\label{fig:4}
\end{figure}

Interestingly enough, as detailed in \cite{PRL13Ruffini}, it turns out that only for an specific range of $\theta_0>0$ these two eigenvalue problems can be solved together, implying as a consequence, a novel reach morphology for the density profiles as well as a novel particle mass bound associated with it. The density profiles presents a quantum degenerate core, followed by a low degenerate plateau until it reach the $r^{-2}$ Boltzmannian regime corresponding to the flat part in the rotation curve. In Fig.~\ref{fig:4} we show a family of density profiles for different values of $\theta_0$ which fulfills the mentioned halo constraints. We also plotted for comparison the purely Boltzmannian curve which agrees with the same \textit{observed} halo magnitudes.

As can be seen from Fig.~\ref{fig:4}, we obtain from this novel analysis a more stringent lower mass bound for the `ino' mass $m$, which is $\sim10$ times higher than the ones inferred in \cite{AA83Ruffini} and \cite{PRL79Tremaine}. This is, $m\geq0.42$ keV$/c^2$ for typical spiral galaxies.

It is interesting that the quantum and relativistic treatment of the configurations considered here are characterized by the presence of central cored structures unlike the typical cuspy configurations obtained from a classic non-relativistic approximation, such as the ones of numerical N-body simulations in \cite{ApJ97Navarro}. This naturally leads to a possible solution to the well-known core-cusp discrepancy \cite{AJ01deBlok}.

\begin{acknowledgments}
We appreciate to Asia Pacific Center for Theoretical Physics (APCTP) for its hospitality during completion of this work.
\end{acknowledgments}


\begin{references}
\bibitem{footnote} `Ino' refers to any beyond standard model neutrino which is suitable within the astrophysical application of the model here presented. For example: the sterile neutrino (see e.g. \cite{ARNPS09Boyarsky}).
\end{references}
\end{document}